\documentclass[aps,pra,reprint, amsmath, amssymb,superscriptaddress, longbibliography]{revtex4-1}
 
\usepackage{bm}
\usepackage[retainorgcmds]{IEEEtrantools}
\usepackage{graphicx}
\usepackage{mathrsfs}
\usepackage{amsmath}
\usepackage{amsfonts}
\usepackage{amssymb}
\usepackage{color}
\usepackage{times,txfonts}
\usepackage{nicefrac}
\usepackage{ragged2e}
\usepackage{tikz}
\usepackage{physics}

\usepackage[colorlinks=true,linkcolor=blue,urlcolor=blue,citecolor=blue,pdfusetitle]{hyperref}

\usepackage{blindtext}

\definecolor{cadmiumgreen}{HTML}{097969}

\begin{document}

\title{Waiting time statistics for a double quantum dot coupled with an optical cavity}
\date{\today}
\author{Luis F. Santos}
\email{luisf@usp.br}
\affiliation{Instituto de F\'isica da Universidade de S\~ao Paulo,  05314-970 S\~ao Paulo, Brazil.}
\author{Gabriel T. Landi}
\email{glandi@ur.rochester.edu}
\affiliation{Instituto de F\'isica da Universidade de S\~ao Paulo,  05314-970 S\~ao Paulo, Brazil.}
\affiliation{Department of Physics and Astronomy, University of Rochester, Rochester, New York 14627, USA}

% \affiliation{Instituto de F\'isica da Universidade de S\~ao Paulo,  05314-970 S\~ao Paulo, Brazil.}

\begin{abstract}
A double quantum dot coupled to an optical cavity is a prototypical example of a non-trivial open quantum system. 
Recent experimental and theoretical studies show that this system is a candidate for single-photon detection in the microwave domain. 
This motivates studies that go beyond just the average current, and also take into account the full counting statistics of photon and electron detections. 
With this in mind, here we provide a detailed analysis of the waiting time statistics of this system within the quantum jump unraveling, which 
% The formalism of open quantum systems has gained increasing popularity in the analysis of physical problems, as it allows us to construct more realistic models by incorporating an external environment that influences the dynamics of the system of interest. Among the myriad of possibilities, two models that have been particularly prominent in recent years are the double quantum dot and the optical cavity. The former is regarded as a fermionic system, permitting the entry and exit of electrons, while the latter exhibits a bosonic nature, enabling the injection and ejection of photons. The coupling of these two entities gives rise to a composite system with diverse empirical applications, particularly in the detection of isolated photons. In light of this, our work aims to introduce the approach of waiting time statistics to study this type of coupling. This formalism 
allows us to extract analytical expressions for the success and failure probabilities, as well as for the inter-detection times.
% Consequently, we have calculated quantities such as the probability of a photon contained in an optical cavity leaking or interacting with the double quantum dot under different scenarios. 
% This  enables us to identify conditions that promote a certain predominance of interaction in the model. 
Furthermore, by comparing single and multi-photon scenarios, we infer a hierarchy of occurrence probabilities for the different events, highlighting the role of photon interference events in the detection probabilities. 
Our results therefore provide a direct illustration of how waiting time statistics can be used to optimize a timely and relevant metrological task. 
\end{abstract}

\maketitle{}

\section{Introduction}

The use of waiting time statistics formalism extends to the analysis of diverse stochastic processes, encompassing realms such as quantum optics \cite{vyas1988waiting, carmichael1989photoelectron}, electronic transport \cite{haack2014distributions, thomas2013electron, dasenbrook2015electron, rajabi2013waiting}, thermodynamic uncertainty relations \cite{friedman2017quantum, garrahan2017simple}, and entropy production estimation \cite{skinner2021estimating}, as pioneered by Stratonovich \cite{stratonovich1967topics}. This powerful tool finds application in quantum master equations, enabling the study of system dynamics through quantum jumps \cite{chia2017hitting, landi2023patterns}, shifting the focus from explicit solutions of the von Neumann equation to the examination of waiting time distributions (WTDs) \cite{brandes2008waiting, landi2024current, albert2011distributions, albert2012electron, brandes2016feedback, kosov2016distribution, ptaszynski2017waiting, schulz2023waiting}. Notably, this approach proves valuable when describing systems with discrete phenomena, such as the punctual creation or annihilation of modes, as is the case with the coupling of a double quantum dot (DQD) to an optical cavity (OC).

The theoretical framework of the DQD-OC coupling is intricate, representing a nontrivial fermionic system interacting with a nontrivial bosonic system, allowing for analytical solutions only under specific considerations \cite{zenelaj2022full}. These solutions unveil the statistical nature and constraints of the system \cite{zenelaj2022full, xu2013full, wong2017quantum, ghirri2020microwave}. Beyond its theoretical richness, the DQD-OC system holds direct experimental relevance in diverse fields, including spectroscopy \cite{oosterkamp1998microwave, michalet2007detectors}, photon observation in astronomy \cite{huber2013observing}, the implementation of quantum circuits \cite{nian2023photon}, and emerging THz quantum technologies \cite{todorov2024thz}. Notably, the system's distinctive feature lies in its ability to function as a photodetector with a unique characteristic: the capacity to capture microwave-scale photons, an energy range four to five times smaller than the optical regime \cite{khan2021efficient}.

By employing the \textcolor{black}{waiting time statistics} formalism, the analysis of the DQD-OC system enables the explicit determination of the probabilities associated with the success or failure of photon detection. Additionally, this approach complements the estimation of quantum efficiency through full counting statistics \cite{zenelaj2022full, xu2013full}, offering valuable insights into the performance of this system in photon detection applications.

In this paper we investigate the waiting time statistics of a DQD-OC setup, unraveled in terms of clicks associated to both electron detection and photon leaks. 
We focus on regimes allowing for analytical solutions, and show that these can shed light on the potential applications of these devices as single-photon detectors. 

\textcolor{black}{This work is structured as follows: In Section \ref{s2}, we provide a detailed description of the DQD and the OC, along with the formalism describing their coupling. Section \ref{s3} introduces the fundamentals of the waiting time statistics formalism, drawing heavily from Section VI.A of \cite{landi2024current}. In Section \ref{s4}, we present our main results, applying the waiting time statistics formalism to analyze the probabilities of photon leakage or absorption in the DQD-OC setup. This analysis is conducted for two distinct initial conditions, with a discussion of their physical significance and relevance. Section \ref{s5} explores how our findings relate to the quantum efficiency of photodetectors based on DQD-OC systems, such as the one studied in \cite{khan2021efficient}. Finally, Section \ref{s6} provides a summary of our results and concluding remarks.}

\section{System and Model} \label{s2}

%The DQD can be conceptualized as two fermionic reservoirs (leads), each coupled to a potential well capable of accommodating a single fermion at any given moment. Within

\begin{figure}
    %\centering
    \includegraphics[scale=0.4]{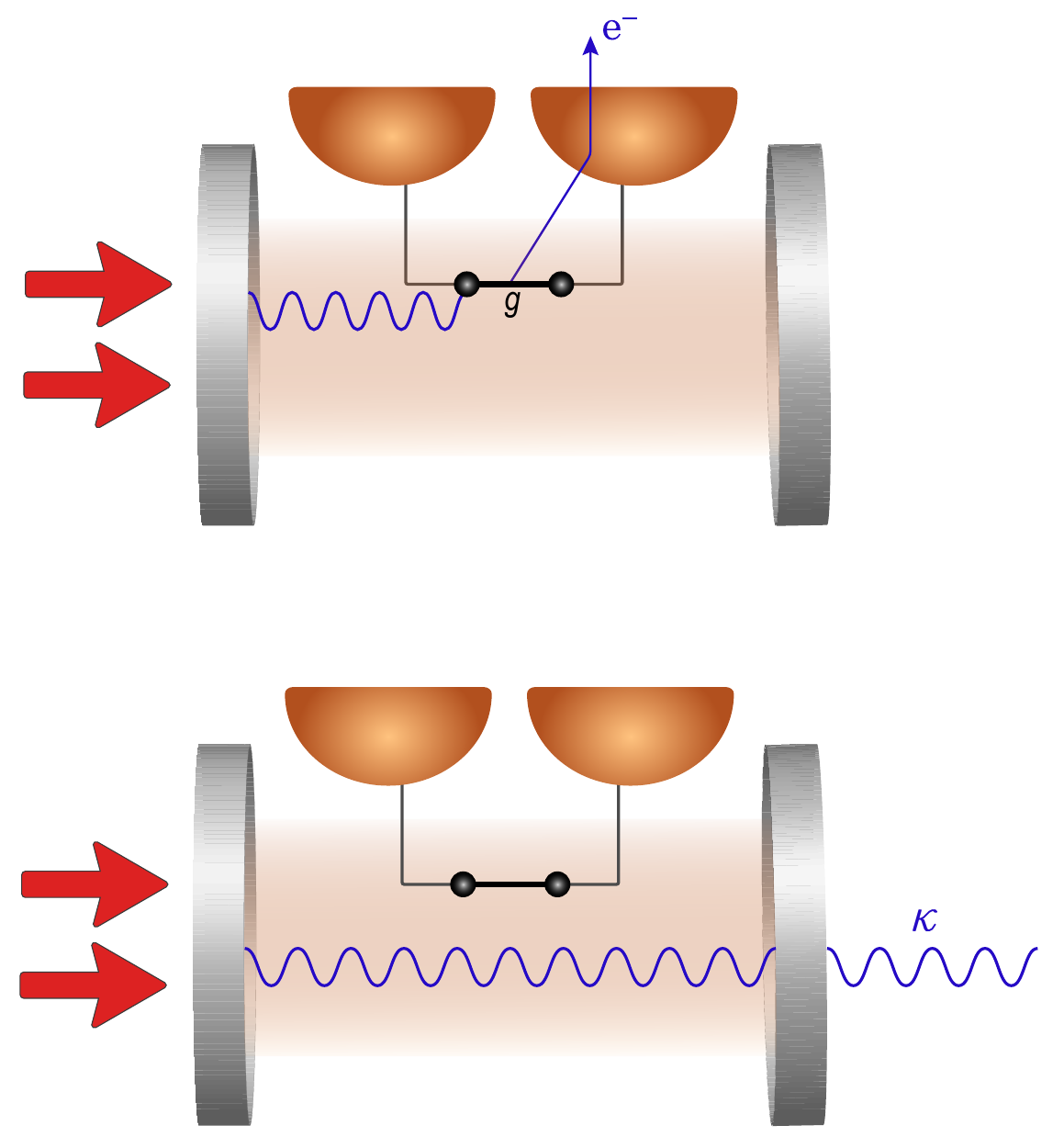}
    \caption{\textcolor{black}{Schematic representation of a double quantum dot (DQD) system coupled to an optical cavity (OC). The DQD consists of two particle reservoirs (depicted in orange) each coupled to a potential well (black spheres). An electron in the ground state can exist in a superposition between the two potential wells. The OC is represented by two gray plates, which receive a photon pump (red arrows) and selectively allow photons of a specific frequency to persist within it. A surviving photon is illustrated by the blue wiggly line.
    The top panel depicts the interaction (with coupling intensity $g$) between the surviving photon in the OC and the electron in the DQD's ground state. In this process, the electron absorbs the photon, transitions to the excited state, and tunnels into the right particle reservoir, generating a detectable photocurrent. The excited electron is represented by the blue arrow.
    The bottom panel illustrates a failed photodetection process: the surviving photon does not interact with the DQD electron and simply passes through the OC, leaking out at a rate $\kappa$.}
    %\textcolor{red}{GTL: eu evitaria falar de cores. Mais facil falar da ``upper wiggly line'' e ``lower wiggly line''. Outra coisa, a blue arrow apontando pra cima nao esta muito visivel. }
    }
    \label{fig dqd-co}
\end{figure}

Our focus lies in the investigation of a system comprising a double quantum dot (DQD) coupled with an optical cavity (OC) \cite{khan2021efficient, xu2013full}, as depicted in Figure \ref{fig dqd-co}. \textcolor{black}{The DQD, highlighted in orange and black in Figure \ref{fig dqd-co}, can be modeled as two connected potential wells, each capable of confining a single fermion at a time. Additionally, each well is coupled to a particle reservoir, typically metallic leads, which can exchange free electrons by either donating or receiving them.} In the Coulomb blockade regime, disregarding electron spin and focusing solely on its presence in the potential wells, the Hamiltonian describing the DQD is given by
\begin{equation}
H_{DQD} = \frac{\epsilon}{2}\left(\ket{R}\bra{R}-\ket{L}\bra{L}\right) + t_c\left(\ket{R}\bra{L} + \ket{L}\bra{R}\right),
\end{equation}
where $\ket{0}$ represents the Fock state denoting the absence of electrons ($Z=0$), the presence of an electron in the right dot ($Z=R$), or in the left dot ($Z=L$). Here, $\epsilon$ signifies the energy of the dots, and $t_c$ represents the coupling energy between them. Employing a transformation matrix
\begin{equation}
\begin{pmatrix}
\ket{L} \\
\ket{R}\\
\ket{0} 
    \end{pmatrix} =  
    \begin{pmatrix}
-\frac{\epsilon}{\Omega} & \frac{2t_c}{\Omega} & 0  \\
-\frac{2t_c}{\Omega} & -\frac{\epsilon}{\Omega} & 0 \\
0 & 0 & 1  
    \end{pmatrix} 
    \begin{pmatrix}
\ket{g} \\
\ket{e}\\
\ket{0} 
    \end{pmatrix}, \quad \Omega \equiv \sqrt{4t_c^2 + \epsilon^2},  
\end{equation}   
we redefine $H_{DQD}$ in terms of the excited ($\ket{e}$) and ground ($\ket{g}$) eigenstates, yielding
\begin{equation}
H_{DQD} = \frac{\Omega}{2}\left(\ket{e}\bra{e} - \ket{g}\bra{g}\right) \equiv \frac{\Omega}{2}\sigma_{3}. \label{hamiltonian DQD}
\end{equation}

%Simultaneously, the OC is envisioned as the combination of a photon pump (e.g., LASER) with a geometric entity that predominantly selects photons possessing a specific frequency of interest 

\textcolor{black}{On the other hand, the OC, represented as the gray plates in Figure \ref{fig dqd-co}, is a specialized cavity that, when subjected to a photon pump (illustrated by the red arrows in the figure, such as a LASER), exhibits geometric and chemical properties that selectively permit, on average, only photons with a specific frequency to enter} \cite{roberts2020driven}. The total Hamiltonian $H_{OC}$ for this system is expressed as
\begin{equation}
H_{OC} = \omega_r a^{\dagger}a + \xi \left(e^{i\omega_lt }a^{\dagger}+e^{-i\omega_l t}a\right), \label{hamiltonian CO}
\end{equation}
where $a$ is a bosonic mode, $\omega_r$ denotes the resonance frequency of the cavity, $\omega_l$ is the pump frequency and $\xi$ is the pump strength. 

Finally, to the lowest order, \textcolor{black}{the DQD-OC coupling can be approximated by the Jaynes-Cummings Hamiltonian} \cite{khan2021efficient}
\begin{equation}
H_I = g\left(a^{\dagger}\sigma_{-} + a \sigma_+\right), \label{H detector}
\end{equation}
where $\sigma_{+} \equiv \ket{e}\bra{g}$ is the raising operator for the DQD  and $\sigma_{-} \equiv \sigma_{+}^{\dagger}$.

The unitary dynamics of the DQD-OC system, expressed by a Hamiltonian $H$ comprising (\ref{hamiltonian DQD}), (\ref{hamiltonian CO}), and (\ref{H detector}), is given in the rotating frame at the pump frequency, as \cite{zenelaj2022full}
\begin{equation}
H = \Delta_d \frac{\sigma_3}{2} + \Delta_r a^{\dagger}a + g\left(a^{\dagger}\sigma_{-} + a \sigma_+\right) + \xi \left(a^{\dagger}+a\right), \label{hamiltonian}
\end{equation}
with $\Delta_d \equiv \Omega - \omega_l$ ($\Delta_r \equiv \omega_r - \omega_l$) representing the difference between the frequency of the DQD (OC) and the frequency of the pump.

Given the weak coupling in the DQD-OC system, the nonunitary dynamics can be modeled using independent dissipators for the DQD and OC~ \cite{breuer2002theory, santos2016microscopic}. Specifically, focusing on single-photon dissipation, $\kappa \mathcal{D}[a]$ becomes the sole dissipator for the open dynamics of the OC, where $\kappa$ quantifies the dissipation rate and $\mathcal{D}[a]\rho = a \rho a^\dagger - \frac{1}{2}\{a^\dagger a,\rho\}$. In the case of the DQD, the ideal photodetector regime \cite{zenelaj2022full, khan2021efficient} is adopted, with interest centered on the input of electrons in the ground state (i.e., $\ket{0}\rightarrow\ket{g}$) and the output of electrons in the excited state (i.e., $\ket{e}\rightarrow\ket{0}$), both occurring at the same rate $\Gamma$. This is captured by $\Gamma \mathcal{D}[s_g^{\dagger}]$ and $\Gamma \mathcal{D}[s_e]$, respectively, with $s_j \equiv \ket{0}\bra{j}$ ($j=g,e$) representing the extraction of an electron in either the ground or excited states, \textcolor{black}{with its Hermitian conjugate expressing the injection of electrons into the corresponding state}.

Consequently, the state $\rho$ governing the DQD-OC system is assumed to follow the Lindblad equation for its open dynamics
\begin{equation}
\Dot{\rho} = -i\left[H,\rho\right] + \Gamma\mathcal{D}[s_g^{\dagger}]\rho + \Gamma\mathcal{D}[s_e]\rho + \kappa \mathcal{D}[a]\rho, \label{dynamic eq}
\end{equation}
where $H$ is given by equation (\ref{hamiltonian}).

\section{Waiting Time Statistics} \label{s3}

Prior to delving into the evaluation of pertinent quantities for DQD-OC system, it is instructive to introduce fundamental concepts of waiting time formalism \cite{landi2024current, brandes2008waiting}. The Lindblad equation (\ref{dynamic eq}) is recast as
\begin{equation}
\dot{\rho}= \mathcal{L}(\rho),
\end{equation}
where $\mathcal{L}(\rho)$, given by the right side of (\ref{dynamic eq}), stands as the Liouvilian operator of the model. This formulation enables the expression of a formal solution
\begin{equation}
\rho (t) = e^{\mathcal{L}t}\rho(0),
\end{equation}
which can be expanded in Dyson's series as
\begin{equation*}
\rho (t) = e^{\mathcal{L}_0t} \rho (0) + \sum_{k\in \mathcal{M}} \int_0^t dt_1 e^{\mathcal{L}_0(t-t_1)}\mathcal{L}_k e^{\mathcal{L}_0 t_1}\rho(0) + \
\end{equation*}
\begin{equation}
+\sum_{k,q\in \mathcal{M}} \int_0^t dt_2 \int_0^{t_2} dt_1 e^{\mathcal{L}_0(t-t_2)}\mathcal{L}_k e^{\mathcal{L}_0 (t_2 - t_1) }\mathcal{L}_q e^{\mathcal{L}_0 t_1}\rho(0) + ... , \label{Dyson}
\end{equation}
where
\begin{equation}
\mathcal{L}_j(\rho)=L_j\rho L_j^{\dagger}, \label{jump super}
\end{equation}
represents the jumps observable in the system, and
\begin{equation}
\mathcal{L}_0 \equiv \mathcal{L} - \sum_{j\in\mathcal{M}} \mathcal{L}_j \label{no jump super}
\end{equation}
is the no-jump operator. Here we also introduce the set $\mathcal{M}$ representing the jump operators which we assume can be monitored. 

Each term in the expansion~\eqref{Dyson} corresponds to the probability associated with a specific number of jumps in the system. Notably, the probability of a jump occurring in the $j$-th channel at any given time is defined as a Waiting Time Distribution (WTD), expressed as
\begin{equation}
W(j,t|\rho) = \Tr{\mathcal{L}_j e^{\mathcal{L}_0t}\rho}
\label{WTD}.
\end{equation}
Marginalizing over $t$, and assuming that the initial state is such that  a jump must necessarily occur, yields 
\begin{equation}
W(j|\rho) = \Tr{\mathcal{L}_j\left(\int_0^{\infty}dte^{\mathcal{L}_0t}\right)\rho} = -\Tr{\mathcal{L}_j\mathcal{L}_0^{-1}\rho}, \label{probabilities independent of time}
\end{equation}
which quantifies the likelihood of a jump in channel $j$, given that the initial state was $\rho$.
Conversely, marginalizing over $j$ yields 
\begin{equation}
    W(t|\rho) = - \Tr\big\{ \mathcal{L}_0 e^{\mathcal{L}_0 t} \rho\big\},
\end{equation}
which is the probability distribution that the first jump occurs at time $t$, irrespective of in which channel it happens. 

Similarly, for scenarios involving two jumps—one at time $t_1$ in channel $j_1$ and another at time $t_2$ in channel $j_2$—the associated probability distribution is given by
\begin{equation}
W \left(j_1,t_1,t_2, j_2|\rho \right)=\Tr{\mathcal{L}_{j_2} e^{\mathcal{L}_0 t_2}\mathcal{L}_{j_1}e^{\mathcal{L}_0 t_1}\rho}. \label{WTD dois pulos}
\end{equation}

Furthermore, these distributions can be employed to define an average time for an event to occur in the system:
\begin{equation}
\langle t \rangle = \int_0^{\infty} dt W(t|\rho) t= -\Tr{\mathcal{L}_0^{-1}\rho}. \label{avarage time}
\end{equation}
This quantity holds significance as it characterizes the characteristic time of the system's evolution, playing a pivotal role in defining quasi-static processes in Thermodynamics \cite{deffner2019quantum}.

\section{Waiting Time Statistics of the DQD-OC system} \label{s4}

The objective of this study is to formulate the probability distributions of success and failure in the detection of a photocurrent, given the presence of a photon within the cavity. The failure process is associated with photon leakage, while the success process is correlated with photon absorption by an electron. To address this problem analytically, \textcolor{black}{we adopt the ideal photodetector regime (as described in Section \ref{s2}) and assume a weak pumping regime, characterized by a low photon influx into the cavity driven by the pump}. The weak pump approximation is introduced by envisioning the activation of the pump, followed by a waiting period until the cavity absorbs $n$ photons. Due to the weak pump, the photon count remains nearly constant during this interval. Consequently, we can analyze the system's dynamics within this time frame, treating the pump as negligible by setting $\xi = 0$ and establishing an initial condition in the density matrix representing the $n$ initial photons in the cavity. We denote different choices of initial conditions as the ``$n$ photon scenario'', specified by the initial density matrix
\begin{equation}
    \rho_0^{(n)} = \ket{\psi_n}\bra{\psi_n}, \quad \ket{\psi_n} \equiv \ket{0}\otimes \ket{n}
\end{equation}
where \(\ket{n}\) is the Fock state of \(n\) photons. 
% Given the Liouvillian \(\mathcal{L}\) governing the dynamics of the DQD-OC system, namely
% \begin{equation}
%     \mathcal{L}\rho = -i\left[H,\rho\right] + \Gamma\mathcal{D}[s_g^{\dagger}]\rho + \Gamma\mathcal{D}[s_e]\rho + \kappa \mathcal{D}[a],
% \end{equation}
% the formal solution to our problem is
% \begin{equation}
% \rho(t) = \exp{{\mathcal{L}t}}\rho_0^{(n)},
% \end{equation}
% which enables us to construct waiting time distributions.

In the first step toward building a waiting time distribution, we identify the channels we can monitor—specifically, both the electron detection ($s_e$) and photon leakage channels ($a$). The channels of interest are the photocurrent channel (right reservoir in Figure \ref{fig dqd-co}) and the photon leak channel (photon accompanied by \(\kappa\) in the same figure). The photocurrent channel can be represented by
\begin{equation}
\mathcal{L}_{e} \rho \equiv  \Gamma s_e \rho s_e^{\dagger}, \label{jump e}
\end{equation}
with jumps \(\ket{e}\rightarrow \ket{0}\) in DQD, occurring at a rate of \(\Gamma\). The photon leak channel is modeled by
\begin{equation}
    \mathcal{L}_{\gamma} \rho \equiv  \kappa a \rho a^{\dagger}, \label{jump a}
\end{equation}
resulting in photon leakage from the cavity to the environment at a rate of \(\kappa\).

Utilizing (\ref{jump e}) and (\ref{jump a}), we define a no-jump Liouvillian, implicitly determining the channels we lack access to:
\begin{equation}
    \mathcal{L}_0 = \mathcal{L} - \kappa\mathcal{L}_{\gamma} - \Gamma \mathcal{L}_e   \label{L0 art}
\end{equation}

This no-jump Liouvillian is employed to evaluate the probabilities of interest in a given scenario.

\subsection*{One Photon Scenario}

We first consider the case \(n=1\). In this scenario, two probabilities are of interest: the probability \(p_{\gamma}\) of a single photon leaking to the environment and the probability \(p_{e}\) of this photon being absorbed by an electron, resulting in a photocurrent. These probabilities are evaluated using Eq.~\eqref{probabilities independent of time}, i.e., 
\begin{equation}
    p_j \equiv W\left(j|\rho\right) = -\Tr{\mathcal{L}_j\mathcal{L}_0^{-1}\rho_0^{(1)}}.
\end{equation}
The assumption that we start with a single photon in the cavity allows us to truncate the bosonic Hilbert space, and therefore obtain the following analytical expression for the success probability: 
\begin{equation}
    p_e = \frac{C}{C+1} \frac{\alpha^2}{(\alpha+1)^2},
    \label{pe n=1}
\end{equation}
where \(C \equiv 4g^2/(\Gamma \kappa)\) is the cooperativity~\cite{aspelmeyer2014cavity, haroche2006exploring} and \(\alpha \equiv \Gamma/\kappa\) quantifies the competition between the electronic and bosonic dissipation rates. 
The failure probability is  $p_{\gamma} = 1 - p_e$.

Notably, as \(C\rightarrow 0\) or \(\alpha \rightarrow 0\), \(p_e \rightarrow 0\), or equivalently, \(p_{\gamma} \rightarrow 1\). This is expected, as \(\alpha \rightarrow 0\) implies more intense interaction of bosonic modes with the environment than fermionic modes, while \(C\rightarrow 0\) indicates weak interaction between DQD and OC compared to their individual interactions with the environment. In both cases, photon absorption by an electron is attenuated. 
% % Conversely, if \(\alpha \rightarrow \infty\), we have
% \begin{equation}
%     \lim_{\alpha \rightarrow \infty} p_e = \frac{C}{(C+1)}, \label{pe alpha large}
% \end{equation}
% and if \(C\rightarrow \infty\),
% \begin{equation}
%     \lim_{C \rightarrow \infty} p_e = \frac{\alpha^2}{(\alpha+1)^2}.
% \end{equation}
Conversely, \(\alpha, C \gg 1\) implies \(p_e \approx 1\), which is reasonable. 
% However, the quadratic dependence on \(\alpha\) terms in the \(C\rightarrow \infty\) regime, compared to the linear dependence on \(C\) in the \(\alpha \rightarrow \infty\) limit, provides insight into the sensitivity of the system to \(\alpha\) versus \(C\). In subsequent sections, we demonstrate that \(\alpha\) serves as a key parameter characterizing the working scenario.

\begin{figure}
    %\centering
    \includegraphics[scale=0.6]{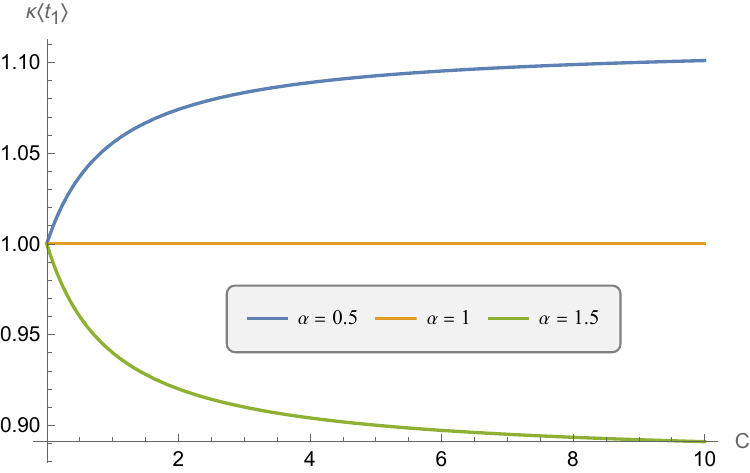}
    \caption{Dimensionless average time for a one photon scenario in terms of $C\in [0,10]$ for different values of $\alpha$. The color scheme represents varying degrees of fermionic-bosonic interaction strength: blue for $\alpha=0.5<1$, orange for $\alpha=1$, and green for $\alpha=1.5>1$. Note the distinct behaviors of $\kappa \langle t_1 \rangle$ across different $\alpha$ regimes. When fermionic interaction predominates in tandem with bosonic interaction ($\alpha > 1$), the average time diminishes with cooperativity, indicative of accelerated system dynamics. Conversely, in the presence of bosonic predominance ($\alpha<1$), the system tends towards greater stability, exhibiting slower dissipation with increasing $C$, until reaching a critical threshold ($\kappa\langle t_1 \rangle \rightarrow 1.11$ in the Figure). Lastly, when bosonic and fermionic interactions exhibit equal intensities ($\alpha=1$), yielding a constant average time, it implies a propensity for recurrent dissipation within fixed time intervals.}  
    \label{fig average time}
\end{figure}

Furthermore, we evaluate the (dimensionless) average time \(\kappa\langle t_1 \rangle\) for any of the jumps (\(e\) or \(\gamma\)) to occur in the system, representing the time until an event takes place (eq. \ref{avarage time}), namely
\begin{equation}
    \kappa \langle t_1 \rangle = -\kappa \Tr{\mathcal{L}_0^{-1}\rho_0^{(1)}} = \frac{(\alpha + 1)^2 + C (3\alpha + 1)}{(\alpha + 1)^2 (C+1)}.
\end{equation}
Figure \ref{fig average time} illustrates the behavior of \(\kappa\langle t_1 \rangle\) in terms of \(C\) for three distinct values of \(\alpha\). For \(0 < \alpha < 1\), the average time to an event increases with \(C\), implying that the system takes longer to transition, within an upper bound given by \(\kappa \langle t \rangle < (3\alpha + 1) / (\alpha + 1)^2\). 
Interestingly, we see that if $\alpha = 1$ (equal dissipation rates for the two channels) we get $\kappa \langle t_1 \rangle = 1$, independent of the cooperativity. 
Notice that this is not true for $p_e, p_\gamma$. 
% For \(\alpha = 1\) or \(\alpha = 0\), the average time remains constant, signifying that in the absence of fermionic interaction or with equal intensities between fermionic and bosonic interaction, the system dissipates, on average, within a fixed time interval. In contrast, if \(\alpha > 1\), the system dissipates progressively faster with increasing \(C\). This crucial observation highlights that although increasing \(\alpha\) might lead to an effective photodetector (\(p_e\rightarrow 1\) as \(\alpha\) grows), doing so carelessly could result in events occurring faster than they can be measured.

\subsection*{Two Photon Scenario}

\begin{figure}
    %\centering
    \includegraphics[scale=0.6]{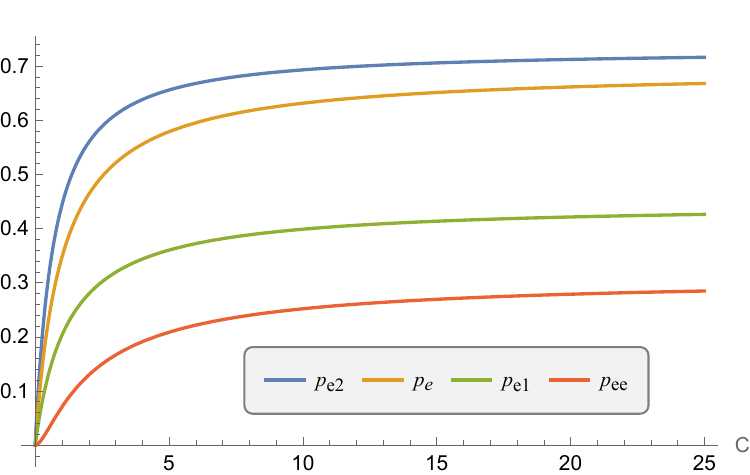}
    \caption{Probabilities of photocurrent detection as a function of cooperativity $C\in[0,25]$, with $\alpha=5$, in a two-photon scenario. The probabilities are color-coded: blue represents $p_{e2}$, the probability of the second detection resulting in a photocurrent; green represents $p_{e1}$, the probability of the first detection resulting in a photocurrent; red represents $p_{ee}$, the probability of both detections resulting in photocurrents. Additionally, in orange, we depict $p_e$, the probability of a photocurrent detection in a single-photon scenario. It is notable that a hierarchical relationship $p_{ee} \leq p_{e1} \leq p_e \leq p_{e2}$ is consistently maintained, with equality observed as $C\rightarrow0$. This observation remains independent of $\alpha$ and provides a reliable means of discerning the scenario under consideration based on the probability distribution of our $n$-th measurement. Conversely, an inverse hierarchy is observed for photon leak probabilities.} 
    \label{fig:photocurrent}
\end{figure}

Next we consider \(n = 2\). In this scenario, four probabilities of interest emerge instead of two: (i) the probability \(p_{ee}\) of both initial photons being sequentially absorbed, resulting in a photocurrent; (ii) the probability \(p_{e\gamma}\) of the first photon being absorbed and the second leaking; (iii) the probability of the first photon leaking, but the second being absorbed; and (iv) the probability \(p_{\gamma\gamma}\) of both photons sequentially leaking. These two-sequential-jump probabilities are defined as (see eq. \ref{WTD dois pulos})
\begin{equation}
    p_{ij} \equiv W(j,i|\rho) = \Tr{\mathcal{L}_j \mathcal{L}_0^{-1}\mathcal{L}_i\mathcal{L}_0^{-1}\rho_0^{(2)}},
\end{equation}
with \(i,j=e,\gamma\). 
Explicitly:
\begin{align}
p_{ee} &= \frac{C^2\alpha^5}{(1+C)(1+\alpha)^2(1+\alpha+C\alpha)(6+5\alpha + \alpha^2)}, \label{p ee}
\\[0.2cm]    
p_{e\gamma} &=   \frac{C \alpha^3\left(C+2C\alpha + (1+\alpha^2)\right)}{(1+C)(1+\alpha)^2(3+\alpha)(1+\alpha + C\alpha)}\label{p e gamma},
\\[0.2cm]    
p_{\gamma e} &= \frac{C\alpha^2 \left(12+\alpha(3+\alpha) (7+\alpha)+C\alpha (9+5\alpha)\right)}{(1+C)(1+\alpha)^2(3+\alpha)(1+\alpha + C\alpha)}, \label{p gamma e}
\\[0.2cm]
    p_{\gamma \gamma}  &= 1- p_{ee}-p_{e\gamma} - p_{\gamma e}.
 % \frac{(1+\alpha)^3(3+\alpha)+C(1+\alpha)(3+\alpha)(1+3\alpha)+C^2\alpha (3+7\alpha)}{(1+C)(1+\alpha)^2(3+\alpha)(1+\alpha + C\alpha)}. \label{p gamma gamma}
\end{align}
% These are properly normalized as \(\sum_{i,j}p_{ij} = 1\).
We can further identify
\begin{equation}
p_{e1} \equiv p_{ee} + p_{e\gamma} = \frac{C\alpha^3}{(2+\alpha)(3+\alpha)(1+\alpha+C\alpha)}    
\end{equation}
and
\begin{align}
    p_{e2} &\equiv p_{ee} + p_{\gamma e}  
    \\[0.2cm]&= \frac{C\alpha^2 \left(12+3(7+3C)\alpha+5(2+C)\alpha^2+(1+C)\alpha^3\right)}{(1+C)(1+\alpha)^2(2+\alpha)(3+\alpha)(1+\alpha+C\alpha)}
\end{align}
as the probability of a jump occurring in the \(e\)-channel (i.e., detection of a photocurrent) in the first and second measurements, respectively. These quantities, along with \(p_{ee}\) (eq. \ref{p ee}) and \(p_{e}\) (eq. \ref{pe n=1}), are plotted in Figure \ref{fig:photocurrent}, where the hierarchy
\begin{equation}
    p_{e2} \geq p_e \geq p_{e1}, \quad \forall \text{ }\alpha, C, \label{hierarchy}
\end{equation}
is observed.
Eq.~(\ref{hierarchy}) indicates that the probability of detecting a photocurrent in the first measurement in the two-photon scenario is lower than in the one-photon scenario, while the opposite holds for the probability of the second measurement resulting in a photocurrent, as it is always greater than the others. This result, independent of \(\alpha\) and \(C\), provides a method for verifying the scenario and highlights the nontrivial interference effects when there are multiple photons inside the cavity.

The asymptotic limits \(\alpha \rightarrow\infty\) and \(C\rightarrow \infty\) yield

\begin{equation}
    \lim_{\alpha\rightarrow \infty }p_{ee} = \frac{C^2}{(1+C)^2},
\end{equation}
and
\begin{equation}
  \lim_{\alpha\rightarrow \infty }p_{e \gamma} =  \lim_{\alpha\rightarrow \infty }p_{\gamma e} = \lim_{\alpha \rightarrow\infty}p_e =\frac{C}{(1+C)^2} \label{limit alpha infty},
\end{equation}
in which we recalled equation (\ref{pe n=1}). This last result indicates that in the strong fermionic interaction regime, the two-photon scenario reduces to a pair of one-photon scenarios, rendering them indistinguishable. However, the same is not true for the large cooperativity regime, where
\begin{equation}
    \lim_{C \rightarrow \infty} p_{ee} = \frac{\alpha^4}{(1+\alpha)^2(6+5\alpha+\alpha^2)},
\end{equation}
\begin{equation}
    \lim_{C \rightarrow \infty} p_{e\gamma} =  \frac{\alpha^2(1+2\alpha)}{(1+\alpha)^2(6+5\alpha + \alpha^2)},
\end{equation}
and
\begin{equation}
    \lim_{C \rightarrow \infty} p_{\gamma e} =  \frac{\alpha^2(9+5\alpha)}{(1+\alpha)^2(6+5\alpha + \alpha^2)}.
\end{equation}
In this case, a nontrivial dependence on \(\alpha\) exists in all cases, preventing specific conclusions. This observation underscores \(\alpha\) as the parameter characterizing the scenarios.
Finally, it is worth noting that
\begin{equation}
    \lim_{\alpha \rightarrow 0} p_{ij} = \lim_{C \rightarrow 0} p_{ij} = \delta_{i\gamma}\delta_{\gamma j},
\end{equation}
which is expected, as previously discussed in the one-photon scenario.

\textcolor{black}{\section{Quantum Efficiency}} \label{s5}

\textcolor{black}{Our work is inspired by the experiment in Ref.~\cite{khan2021efficient}. However, the setup there consists of a steady external photon pump and hence a steady current of photo-electrons. 
This allows the authors to evaluated the quantum detection efficiency, defined as \cite{wong2017quantum}
\begin{equation}
    \eta = \frac{\text{photoelectron count}}{\text{incident photons}},
\end{equation}
evaluated in steady-state. 
Here we are considering instead the scenario where there is no continuous photon pump, but just individual photon injections. This approach therefore does not apply. 
To circumvent this issue, we propose an alternative approach by assuming that the efficiency is proportional to the probability of an electron being converted into a photocurrent. This assumption seems reasonable, as it is intuitive that a higher probability would correspond to a greater conversion rate. Thus, we suggest}
\begin{equation}
\eta \sim p_e \label{ansatz}.
\end{equation}
\textcolor{black}{
To corroborate our definition, we now show that this is indeed related to the steady-state efficiency. 
To do that, we start with Eq.~(31) \textcolor{black}{of} Ref.~\cite{wong2017quantum}, under the assumption that \(\gamma_{1}=\gamma_{2} = 0\) (meaning there is no interaction between the electrons of the DQD and the phonons of the reservoirs).
We then find that }
\begin{equation}
    \eta = \frac{4\kappa g^2 \Gamma \epsilon}{\Omega \left[4\Delta_d^2+\Gamma^2\right] \left[\left(\Delta_r - \frac{4g^2\Delta_d}{4\Delta_d^2 +\Gamma^2}\right)^2 + \left(\frac{\kappa}{2}+\frac{2g^2 \Gamma}{4\Delta_d ^2 + \Gamma^2 }\right)^2 \right]}
\end{equation}
\textcolor{black}{In the resonant case \((\Delta_d = \Delta_r = 0)\), this reduces to}
\begin{equation*}
    \eta =\frac{4\epsilon}{\Omega}\frac{C}{\left(1+C\right)^2 } = 
\end{equation*}
\begin{equation}
    \textcolor{black}{=\frac{4\epsilon}{\Omega}\lim_{\alpha\rightarrow \infty }p_{e \gamma} =  \frac{4\epsilon}{\Omega}\lim_{\alpha\rightarrow \infty }p_{\gamma e} =\frac{4\epsilon}{\Omega}\lim_{\alpha \rightarrow\infty}p_e,} \label{efficiency}
\end{equation}
\textcolor{black}{in which we used Eq. (\ref{limit alpha infty}).} This provides the suggested connection. Namely, the steady-state quantum efficiency studied in ~\cite{khan2021efficient} is directly proportional to single-photon detection probability $p_e$. \textcolor{black}{Furthermore, we observe that the quantum efficiency discussed can be associated with the $\alpha \rightarrow \infty$ regime in the two-photon scenario, providing insight into its potential connection with the $n$-photon scenario.}

\textcolor{black}{Here, we emphasize that Eq. (\ref{efficiency}) serves more as a proof of principle than as a result directly applicable to laboratory experiments. We demonstrate that waiting time statistics can be used to evaluate quantum efficiency, offering the advantage of not requiring the assumption of a steady-state system to derive analytical expressions. In this work, we consider specific initial conditions based on certain physical approximations. Compared with the full counting statistics approach \cite{khan2021efficient, zenelaj2022full}, our approach provides an alternative and complementary method to assess the quality of the DQD-OC coupling within the context of quantum circuits.}

% \textcolor{black}{(see eq. \ref{limit alpha infty}). Comparing the above expression with (\ref{ansatz}), we find that}

% \begin{equation}
%     \eta_1 = \frac{4\varepsilon}{\Omega}p_e
% \end{equation}

% \textcolor{black}{which represents a more general form of quantum efficiency that accounts for the competition between electronic and bosonic modes (encompassed by 
% $\alpha$). This shows that the approximations made in \cite{wong2017quantum} result in a quantum efficiency that is a particular case of our more generalized expression for $n=1$.} 

%\textcolor{red}{GTL: sugiro tomar cuiudado para ver se todos os simbolos que aparecem na Eq. (42) estao definidos corretamente.}

\section{Concluding Remarks} \label{s6}

In conclusion, under the assumption of a weak pump regime, we have leveraged the formalism of waiting statistics to derive probabilities governing the success and failure of photocurrent conversion within a DQD-OC system, examining scenarios involving one and two incident photons. 
\textcolor{black}{The main advantage of our approach is that it yields a time-resolved picture of individual photo-detections, which is much richer than previous approaches that are based only on steady-state currents under a continuous pump.  
With our approach, individual electron detection probabilities become analytically calculable. 
And, similarly, we can evaluate the average response time of the detection, as well as its higher order statistics.}
While the extension of this approach to scenarios involving $n$ photons is conceptually straightforward, it is imperative to note that the validity of this approximation diminishes as $n$ increases, as it fails to account for mixed states at its core, leading to nonphysical outcomes.
Nevertheless, our methodology adequately captures the interference effects between photons within the cavity, significantly influencing the photocurrent detection process. 
% A logical progression involves constructing WTDs for non-zero detunings ($\xi\neq0$), corresponding to scenarios with reasonable to strong pumping. This analysis is anticipated to shed light on how the probabilities of photocurrent conversion evolve with varying pump intensities.

A logical progression involves incorporating additional complexities into our model, such as losses through phononic channels, as outlined in the work by Zenelaj et al. \cite{zenelaj2022full}. This enhancement will contribute to a more realistic representation of the DQD-OC system, accounting for factors beyond the weak-pump approximation and further refining our understanding of the underlying physical processes.

\section*{Acknowledgments}

The authors would like to thank Amanda Candido Ferreira for her kind assistance in creating Figure \ref{fig dqd-co} of this paper. L F S acknowledges the financial support of Coordenação de Aperfeiçoamento de Pessoal de Nível
Superior (CAPES) – Brazil, Finance Code 001.

\bibliography{References}

\end{document}